\def\aeti{$\alpha$-(BE\-DT\--TTF)$_2$\-I$_{3}$}
\def\cm{cm$^{-1}$}
\documentclass[aps,prb,twocolumn,showpacs]{revtex4}
\usepackage{graphicx}%
\usepackage{graphicx}%
\usepackage{dcolumn}%
\usepackage{bm}%
\usepackage{xcolor}

\begin{document}

\title{Electrically-induced phase transition in $\alpha$-(BEDT-TTF)$_2$I$_3$:\\
Indications for high-mobility hot charge carriers}
\author{T. Peterseim}
\author{T. Ivek}
\author{D. Schweitzer}
\author{M. Dressel}
\affiliation{1. Physikalisches Institut, Universit\"at Stuttgart, Pfaffenwaldring 57, 70550 Stuttgart, Germany}

\date{\today}

\begin{abstract}
The two-dimensional organic conductor $\alpha$-(BEDT-TTF)$_2$I$_3$ undergoes a metal-insulator transition at $T_{\rm CO}=135$~K due to electronic charge ordering. We have conducted time-resolved investigations of its electronic properties in order to explore the field- and temperature-dependent dynamics.
At a certain threshold field, the system switches from low-conducting to a high-conducting state,
accompanied by a negative differential resistance. Our time-dependent infrared investigations
indicate that close to $T_{\rm CO}$ the strong electric field pushes the crystal into a metallic state with optical properties similar to the one for $T>T_{\rm CO}$. Well into the insulating state, however, at $T=80$~K, the spectral response evidences a completely different electronically-induced high-conducting state. Applying a two-state model of hot electrons explains the observations by excitation of charge carriers with a high mobility. They resemble the Dirac-like charge-carriers with a linear dispersion of the electronic bands found in $\alpha$-(BEDT-TTF)$_2$I$_3$ at high-pressure.
Extensive numerical simulations quantitatively reproduce our experimental findings in all details.
\end{abstract}


\maketitle

\section{Introduction}
\label{sec:introduction}
The charge-transfer salt \aeti\ is probably the best studied model compound of a metal-in\-sula\-tor transition
in two-dimensional electron systems\cite{Bender84a,Dressel94} that is supposed to be driven by electronic charge order.\cite{Kino95,Kino96,Seo04}
From NMR\cite{Takano01,Takahashi06} and optical spectroscopy,\cite{Moldenhauer93,Wojciechowski03,Dressel04,Yue10,Ivek11,Yakushi12} as well as x-ray diffraction studies\cite{Kakiuchi07} it is well known that below $T_{\rm CO}=135$~K charge disproportionation develops on the BEDT-TTF molecules.
Since its discovery three decades ago, the intererst in the title compound never faded because it exhibits a rich temperature-pressure phase diagram,
with a number of intriguing quantum phenomena ranging from electronic ferroelectricity\cite{Ivek10,Tomic15,Lunkenheimer15}
to superconductivity,\cite{Tajima02,Kobayashi04}
from nonlinear transport\cite{Dressel95,Ivek12} to zero-gap semiconductivity\cite{Tajima06,Mori10}
characterized by Dirac cones and massless Dirac fermions,\cite{Tajima07,Suzumura12,Kajita14,Beyer16}
but also the appearance of persistent photoconduction,\cite{Tajima05} photoinduced phase transition,\cite{Iwai07,Kawakami10,Iwai12} and nonlinear ultrafast optical response.\cite{Yamamoto08}

\begin{figure}[h]
    \centering
       \includegraphics[width=1\columnwidth]{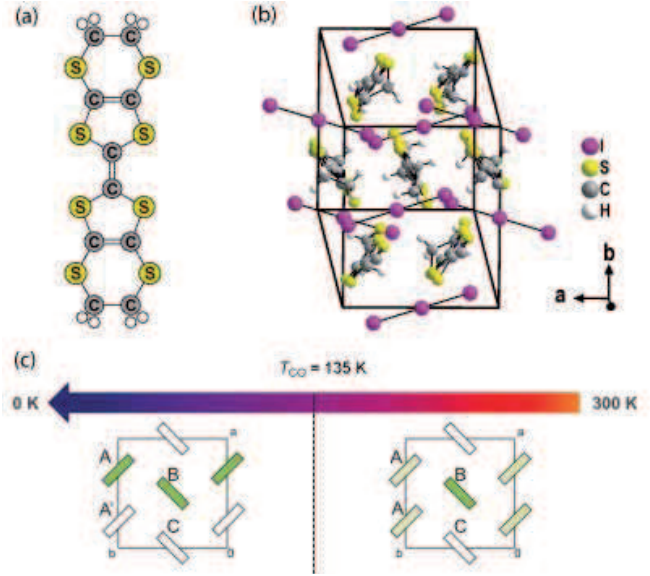}
    \caption{(Color online) (a)~Sketch of the bis-(ethyl\-ene\-di\-thio)\-te\-tra\-thia\-ful\-va\-lene
molecule, called BEDT-TTF. (b)~Drawing of the \aeti\ unit cell with orientation along the $c^{\prime}$-direction. The unit cell contains four BEDT-TTF and two I$_3$ molecules. The cations are arranged in a herringbone-like structure within the $ab$-plane separated from each other along the $c^{\prime}$-direction by the anions.
(c)~Illustration of the development of the charge disproportionation and crystal symmetry in \aeti\ when cooling from room temperature to 0~K. The $ab$-plane is displayed with the four BEDT-TTF molecules labeled by A, B, and C;
the anions are not included for clarity reasons. The A molecules are related to each other by a point inversion on their connecting line. The opacity of the green color represents the charge imbalance between the molecules. At $T=300$~K there is an imbalance between molecule B and C. Below $T_{\rm CO}$ the point inversion is lost and the A molecules become inequivalent. A stripe-like charge-order pattern develops along the $b$-direction.
    \label{fig:structure}}
\end{figure}
At room temperature \aeti\ consists of layers formed by the BEDT-TTF molecules that are separated along the $c^{\prime}$-direction by I$_3^-$ anions creating an electronically two-dimensional system, as displayed in Fig.~\ref{fig:structure}.
At $T_{\rm CO}=135$~K  the symmetry class of the crystal changes from P$\overline{1}$ to P1 at low temperatures, thus breaking the inversion symmetry between the molecules A and A$^{\prime}$.
Now, all four molecules are crystallographically independent and each molecule in the unit cell carries a distinct charge different from the others.
As shown in Fig.~\ref{fig:structure}(c),
the charge order forms horizontal stripes with molecule B still being charge-rich and molecule C charge-poor, but either molecule A or A$^{\prime}$ being charge poor/rich.
The charge ordering leads to a ferroelectric state which was probed by second harmonic generation of light\cite{Yamamoto08,Yamamoto10,Yamamoto12} as a direct proof of the loss of the inversion symmetry.
These studies have also shown that different ferroelectric
\clearpage
\noindent
domains exist in a
single crystal separated from each other by domain walls. Indications of an influence of cooling rate and pressure on the amount of domains and their size was found, although a systematic study is still missing and necessary.

Recent dielectric measurements\cite{Ivek10,Ivek11} detect two modes with a dielectric strength $\Delta\epsilon$ between 400 and 5000 for the $a$- and $b$-direction indicating a strong ferroelectric response with no temperature dependence below $T_{\rm CO}$. The first mode is present at all temperatures and is ascribed to a charge-density wave, which is screened by activated charge carriers since the relaxation time follows the temperature-dependence of the dc conductivity. The second mode appears on further cooling below $T=75$~K. It is suggested that this response is caused by solitons and domain walls which are fixed to pinning centers.
Alternatively the  existence  of  polar  and  nonpolar  stacks  of  the  BEDT-TTF molecules  was suggested,\cite{Lunkenheimer15}  preventing  long-range ferroelectricity.

In a complementary approach, the electrodynamical properties of \aeti\ at the metal-insu\-la\-tor phase transition have been explored by non-lin\-e\-ar and time-dependent studies.\cite{Naito10b,Iimori14} Particular attention was drawn by the observation of photo-in\-duc\-ed phase transition phenomena,\cite{Iwai07,Nakaya08,Kawakami10} photo\-con\-ductivity\cite{Tajima05,Iimori09} connected with a memory effect,\cite{Iimori07,Iimori07b}  non-lin\-e\-ar con\-ductivity\cite{Dressel94,Tamura10,Ivek12} and zero-gap states under pres\-sure.\cite{Katayama06,Mori10,Alemany12,Beyer16} Here we focus on the electrically-in\-duced phase transition in \aeti\
by investigating the time-resolved transport and optical properties. We then apply a model of non-equilibrium charge carriers to our observations explaining our findings in all details.

\section{Experimental Results}
The time-, temperature- and field-dependent transport and optical studies have been conducted
on \aeti\ single crystals, as grown by electrochemical methods.\cite{Bender84a,Bender84b}
The experimental details are given in the Supplemental Materials.\cite{SM}

\begin{figure}
    \centering
    \includegraphics[width=1\columnwidth]{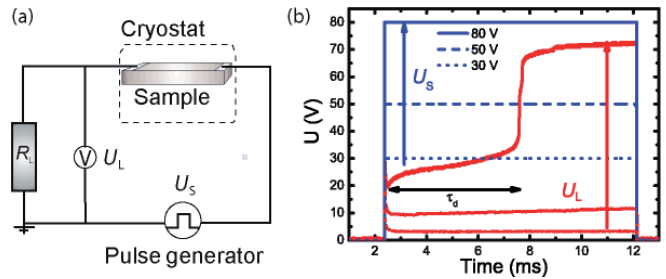}
    \caption{(a)~For time-dependent transport studies a voltage pulse $U_S$ of some milliseconds duration is applied; the voltage drop $U_{L}$ across a load resistor measures the total current $I_{\rm tot}=U_{L}/R_{L}$. (b)~Time-dependence of the source voltage $U_S$ (blue lines) and drop $U_L$ across the load resistor (red curves) measured at $T=80$~K along the $a$-direction of \aeti.  After a short switching pulse,
    a steady state regime is reached for low voltages (dashed and dotted lines). Above a certain threshold field, however, the sample switches from a low-conducting, insulating state into a high-conducting state which is marked by a steep increase of the current.
    \label{fig:SetupPulse}
    }
\end{figure}
Besides conventional dc transport measurements in four-point geometry (cf.\ Figs.~S1 and S2),
dynamical resistivity studies have been performed by applying a voltage of $U_S$ across the single crystal of \aeti\ for a few milliseconds and measuring the total current flow $I_{\rm tot}$ as a function of time. In the metallic but also well into the insulating state, the response is basically constant at low voltage, indicating a steady state. In Fig.~\ref{fig:SetupPulse}(b) it can be seen, however, that for $U_S=80$~V applied at low temperatures ($T=80$~K, for instance), heating continuously increases the current until the sample suddenly turns metallic at certain delay time $\tau_d$. In a first step (Sec.~\ref{sec:transport_studies}) we want to characterize, how this switching evolves with time and how the behavior depends on temperature and electric field applied. In Sec.~\ref{sec:optical_studies} we will then present optical investigations which elucidate the nature of the electrically-induced metallic state.

\subsection{Transport Studies}
\label{sec:transport_studies}
\begin{figure}
    \centering
       \includegraphics[width=0.8\columnwidth]{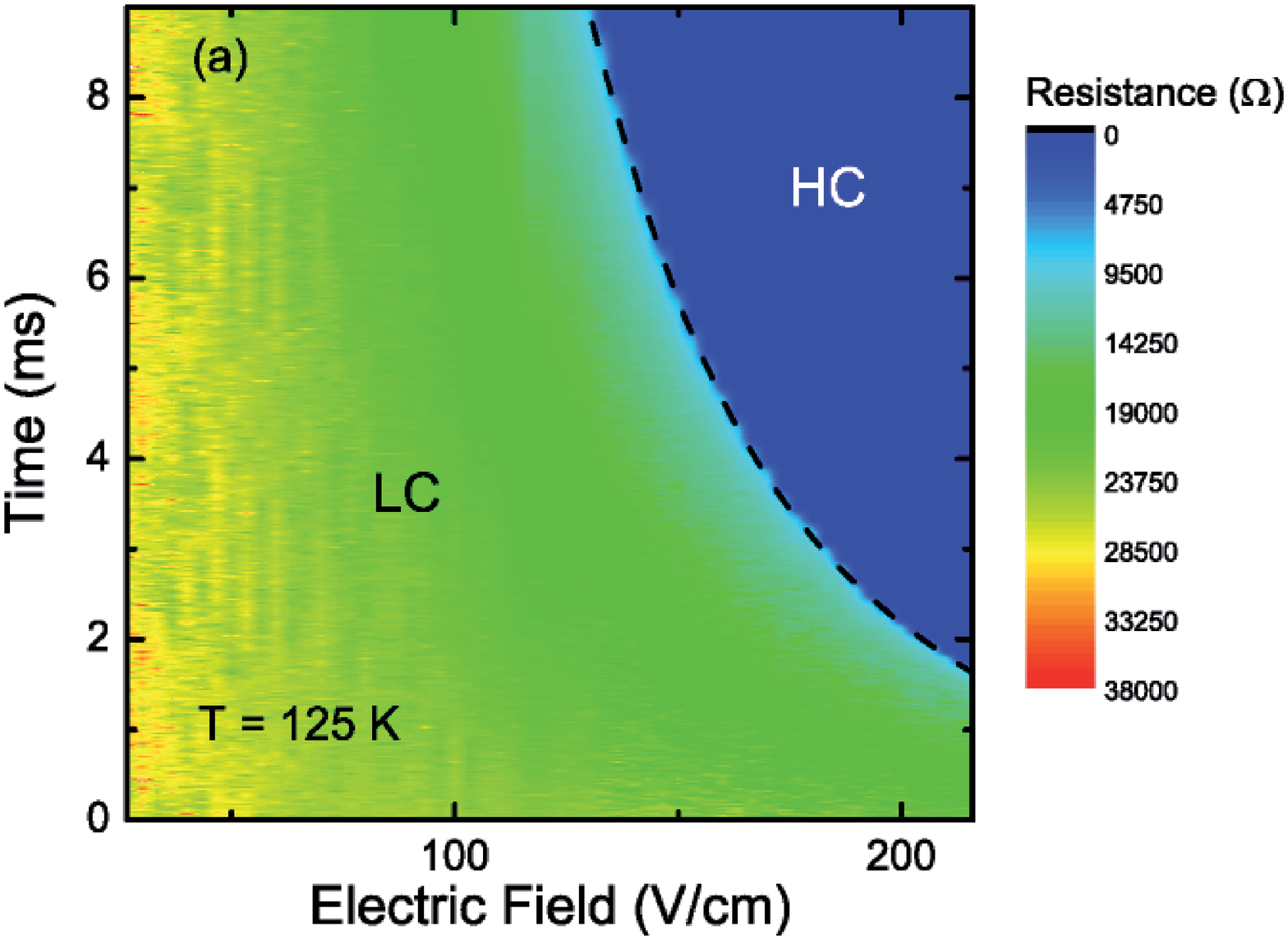}\vspace*{3mm}
       \includegraphics[width=0.8\columnwidth]{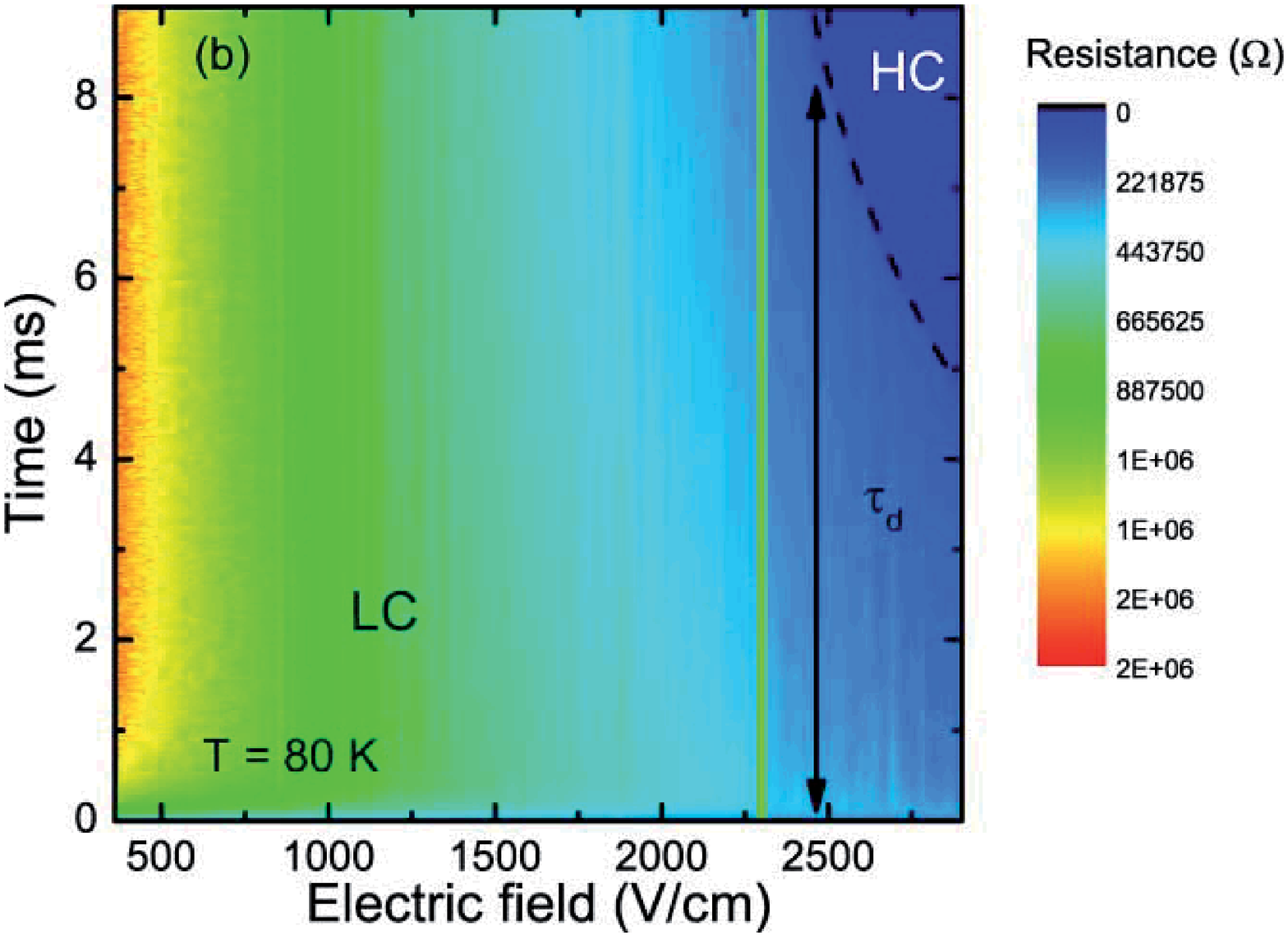}
    \caption{(Color online) Contour plot of the time-dependent sample resistance of \aeti\ in a time frame of 10~ms for different electric fields $E_{\rm sample}$ recorded at (a)~a temperature close to the charge-order transition, $T=125$~K, and (b)~well below, $T=80$~K. The blue area corresponds to a metallic behavior (high conducting, HC), while the green to red areas indicate the insulating state (low conducting, LC).  With increasing field the delay time $\tau_d$ to enter the metallic state decreases. Note that in both frames different scales have been used for the applied field and for the observed resistance, respectively.
    \label{fig:ContourResistance}}
\end{figure}

In Fig.~\ref{fig:ContourResistance} the time dependence of the sample resistance is plotted as a function of the applied electric field for two selected temperatures.
At $T=125$~K the sample resistance $R_{\rm sample}$ is approximately 19~k$\Omega$.
In time frame under inspection it jumps abruptly from the insulating state into a high-conducting state above a certain threshold field that is marked by the color change from green to blue in Fig.~\ref{fig:ContourResistance}.
The resistance is reduced by two orders of magnitude to $R_{\rm sample}\approx 200~\Omega$.
For the contour plot it becomes clear that the switching takes place with a certain delay time $\tau_d$, which decreases with increasing electric field $E_{\rm sample}$ in a nonlinear manner.

For lower temperatures, $T=80$~K, the behavior is rather similar;
however, the total resistance $R_{\rm sample}\approx 1~{\rm M}\Omega$ is much higher and decreases slightly  with increasing voltage due to the nonlinear behavior of the contact resistance.\cite{Ivek12}
Note that at reduced temperatures the electric field must be significantly increased to induce the high conducting state. It should also be pointed out that despite the different absolute values of resistivity recorded, the step height between low- and high-conducting range is again about two orders of magnitude, as in the case of $T=125$~K.

\begin{figure}
    \centering
       \includegraphics[width=0.8\columnwidth]{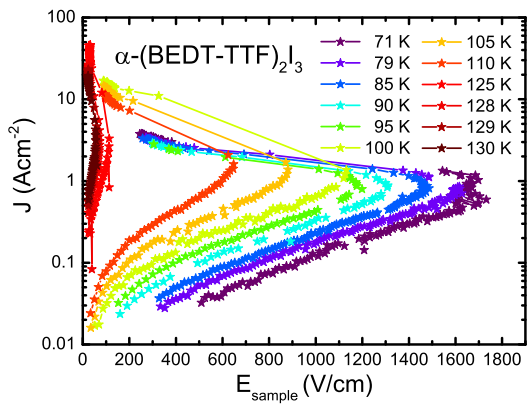}
    \caption{(Color online) Current density $J$ as a function of the electric field $E_{\rm sample}$ recorded at a \aeti\ single crystal at different temperatures between $T=130$ and 80~K as indicated. At all temperatures a negative differential regime is observed, where the threshold electric field increases with decreasing temperature. \label{fig:Current_log}}
\end{figure}
The calculated current-voltage characteristic, $J$ {\it vs.} $E_{\rm sample}$, is displayed in Fig.~\ref{fig:Current_log} for the temperature range from $T=70$ to 130~K.
The $J$ curves exhibit a regime of a negative differential resistance (NDR),
where the current jumps to a much larger value, while the electric field at the sample $E_{\rm sample}$ drops significantly. This curve reveals a so-called S-shape (current-controlled), in contrast to an N-curve (voltage-controlled)
where the current drops to smaller values with increasing electric field due to a reduction of the carrier mobility.
The NDR appears above a certain threshold field $E_{\rm thresh}$, which increases with lowering $T$. Before the NDR regime the current density increases linearly reflecting ohmic behavior. In our context, this means that the conductivity must be significantly enhanced by generating high-mobility carriers.

\begin{figure}
    \centering
       \includegraphics[width=1\columnwidth]{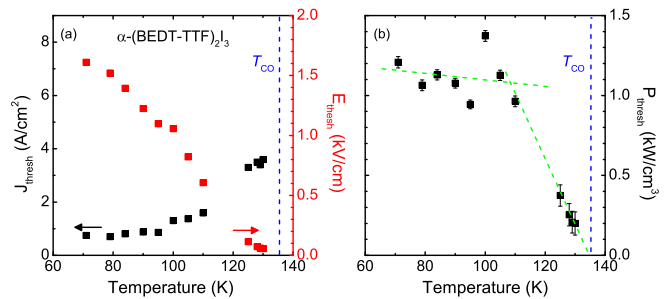}
    \caption{(Color online) (a)~Temperature dependence of the threshold current density $J_{\rm thesh}$ (black, left axis) and threshold electric field $E_{\rm thesh}$ (red, right axis).
    (b)~Power density $P_{\rm thesh}$ below the transition temperature $T_{\rm CO}$ of \aeti.
    The green dotted lines are guides to the eye. The blue dashed vertical lines mark the metal-insulator transition temperature $T_{\rm CO}$.\label{fig:Threshold_exp}}
\end{figure}

From Fig.~\ref{fig:Current_log}, we extract the threshold electric field $E_{\rm thresh}$ and the threshold current density $J_{\rm thesh}$ and plot the parameter as a function of temperature in Fig.~\ref{fig:Threshold_exp}(a). $E_{\rm thesh}$ decreases linearly with $T$ up to $T_{\rm CO}$ where it vanishes, as expected; there is no NDR observed in the metallic state above $T_{\rm CO}$.
This agrees with measurements in the current mode performed in \aeti\ below 100~K;\cite{Tamura10,Itose13} in other organic materials similar observations have been reported.\cite{Ozawa09} There, the threshold field $E_{\rm thesh}$ also rises continuously as $T$ is reduced.

Let us compare the absolute value of the NDR threshold fields found in our experiments to other systems. In the case of typical charge-density-wave materials,\cite{Gruner88} threshold fields
below 1~V/cm are typically observed, while the electric field values we extract from \aeti\ are several orders of magnitude larger. On the other hand, avalanche effects or dielectric breakdown are found in semiconductors or insulators only for fields exceeding 1 MV/cm.
But also the observed temperature dependence is distinct from those phenomena.
For a charge-density-wave the threshold-field follows a $E_{\rm thresh}\propto \exp\left\{-T/T_0\right\}$ behavior,\cite{Maki86} in contrast to dielectric breakdown where it behaves like  $E_{\rm thesh}\propto \exp\left\{T_0/T\right\}$.
The linear $T$ dependence observed here in \aeti\ does not follow either of them.
Hence we conclude that the nonlinear transport behavior in charge-ordered materials plays a special role and cannot be simply classified within the previous mentioned phenomena.

In contrast to $E_{\rm thresh}$, the current density $J_{\rm thresh}(T)$ does not
follow a linear temperature behavior; it seems to diverge towards $T_{\rm CO}$ and approaches a constant
value at low $T$. Similar results have been reported previously.\cite{Itose13}
From both quantities, the power density at the threshold, $P_{\rm thresh}= E_{\rm thresh} \cdot J_{\rm thresh}$, can be readily calculated [Fig.~\ref{fig:Threshold_exp}(b)]. Interestingly, the power necessary to the switch the conduction state rises steeply as the temperature drops below $T_{\rm CO}$;
only for $T < 120$~K  $P_{\rm thresh}$ remains basically constant. This temperature dependence resembles the behavior of the resistivity plotted in Fig.~S2,
where the largest modification occurs in a narrow temperature range $120~{\rm K}< T <135$~K where the energy gap opens.

In order to ensure that this effect is not caused by Joule heating, we have conducted a thorough calculation of the stored energy and the possible temperature rise; details are given in the Supplemental Materials.\cite{SM} Even under the assumption that 100\%\ of the electrical power is used for heating the crystal, the experiments performed at $T=80$~K will never lead to sample temperatures above 96~K. At least far below $T_{\rm CO}=135$~K, we can therefore exclude that the switching from the low-conducting to high conducting state is caused by Joule heating.

\subsection{Optical Studies}
\label{sec:optical_studies}
 \begin{figure}[b]
    \centering
       \includegraphics[width=1\columnwidth]{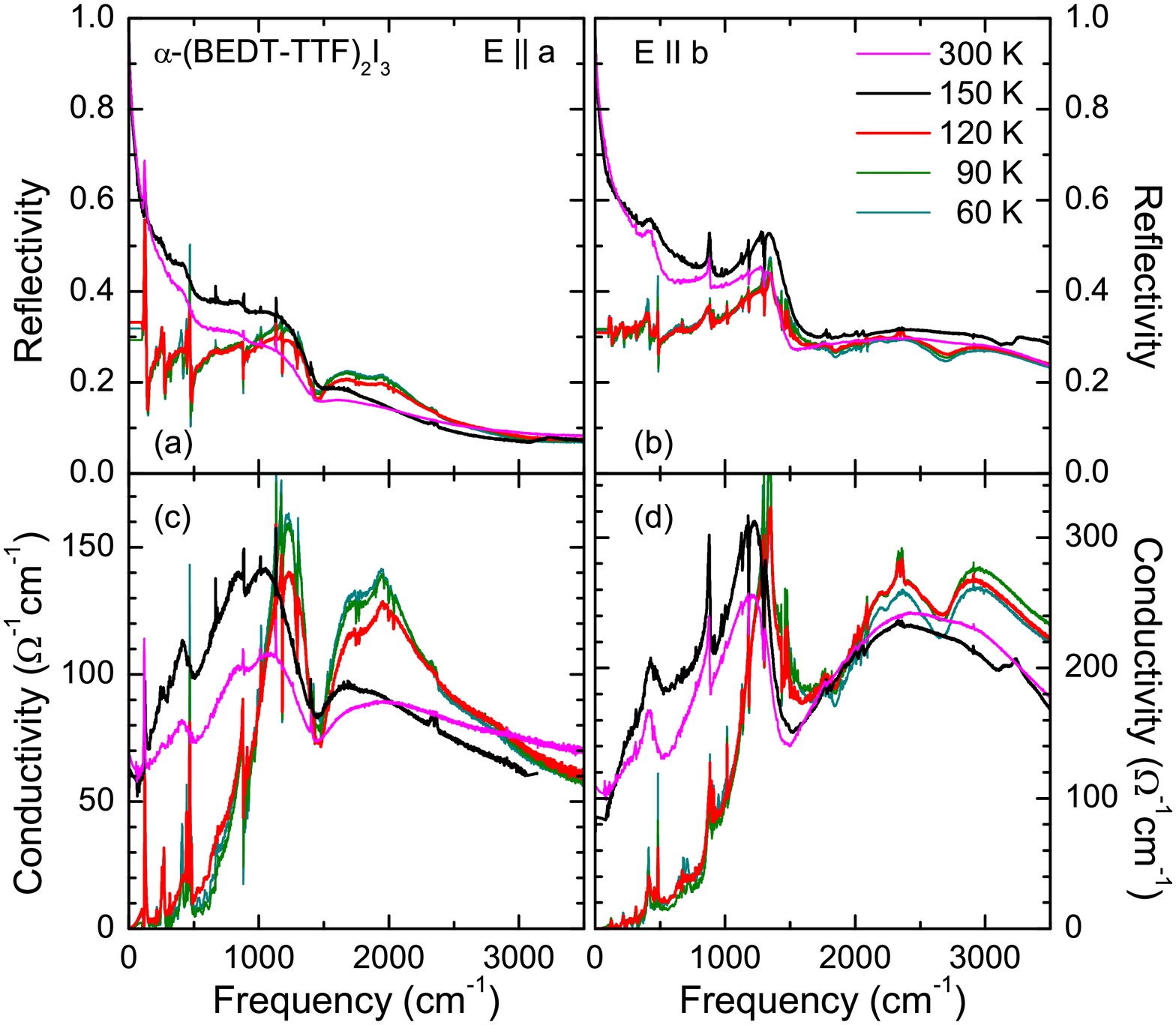}
    \caption{(Color online) Optical properties of \aeti\ for different temperatures as indicated. The upper panels (a) and (b) show the
reflectivity for the polarization of the electric field $E\parallel a$ and
$E\parallel b$. The corresponding optical conductivity is plotted in
the lower panels (c) and (d); note the different vertical scales.
    \label{fig:OptRefCond}}
\end{figure}
Beyond these time-dependent transport studies,
more insight into the nature of the electronic state
can be obtained from the frequency dependent conductivity.\cite{DresselGruner02} It is well known that the metal-insulator transition of \aeti\ at $T_{\rm CO}=135$~K has a strong influence on the optical response probed by polarized infrared reflection measurements. As illustrated in Fig.~\ref{fig:OptRefCond}, the reflectivity decreases below  approximately 1500~\cm\ when the temperature drops below $T_{\rm CO}$, while $R(\omega)$ increases in the mid-infrared region. The corresponding optical conductivity is plotted in Fig.~\ref{fig:OptRefCond}(c) and (d), demonstrating the shift in spectral weight from the frequency range below the gap at approximately 600~\cm\ to the mid-infrared peak around 200~\cm.\cite{Clauss10} The change takes place within a rather narrow temperature interval between $T=140$ and 120~K with basically not variation below.

\begin{figure}
    \centering
       \includegraphics[width=1\columnwidth]{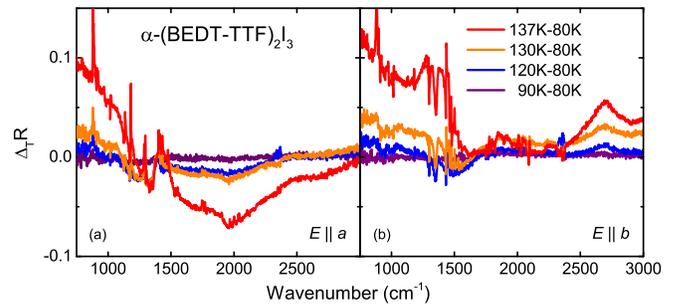}
    \caption{Difference of the optical reflectivity at $T=137$~K (red), 130~K (orange), 120~K (blue), and 90~K (purple) with respect to the reflectivity at$T=80$~K. Panel (a) and (b) correspond to the polarizations $E \parallel a$ and $E \parallel b$, respectively.  \label{fig:DeltaRT}}
\end{figure}

Since we are interested who the reflectivity changes in certain spectral regions, in Fig.~\ref{fig:DeltaRT} we plot the difference in reflectivity $\Delta_T R(\nu)$ between various temperatures above and below $T_{\rm CO}$.  When comparing $T=120$ and 80~K, for instance,
$\Delta_T R$ exhibits almost no modification besides a slight shift of the energy gap in the charge ordered phase. Pronounced spectral changes, however, are found between the two phases, i.e.\ looking at $T=137$ and 80~K. When entering the charge-ordered state below $T_{\rm CO}$ the low-frequency reflectivity rises on the expense of the mid-infrared region. No significant differences are observed between the two polarizations $E \parallel a$ and $E \parallel b$.

In order to characterize the electric-field-induced me\-tal\-lic state spectroscopically and to draw a more complete picture of the dynamics at the metal-insulator transition of \aeti, we have performed time-resolved infrared measurements by employing step-scan Fourier-transform infrared spectrometer Bruker Vertex 80v; for technical details see Ref.~\onlinecite{Peterseim16c}. The time resolution was set to 20~$\mu$s and 1000 time steps were recorded for 20~ms. The data were acquired 1~ms before the voltage pulse was initialized. To improve the signal-to-noise ration, we averaged over 20 scans. The spectral resolution was 4~\cm\ and the recorded frequency ranged from $\nu=800$ to 4000~\cm.
The time-resolved measurements were performed in the ac-mode, highlighting the changes in reflectivity $\Delta_t R(\nu,t)$ as time $t$ elapses, where $\Delta_t R(\nu,t)= R(\nu,t) - R(\nu,0)$. It is of interest to compare this with the change of reflectivity when the temperature $T$ is varied with respect to a reference temperature $T_0$: $\Delta_T R(\nu,T) = R(\nu,T) - R(\nu,T_0)$, as plotted in Fig.~\ref{fig:DeltaRT}.

\subsubsection{Close to metal-insulator transition: $T=125$~{\rm K}}
\begin{figure}[b]
    \centering
       \includegraphics[width=0.8\columnwidth]{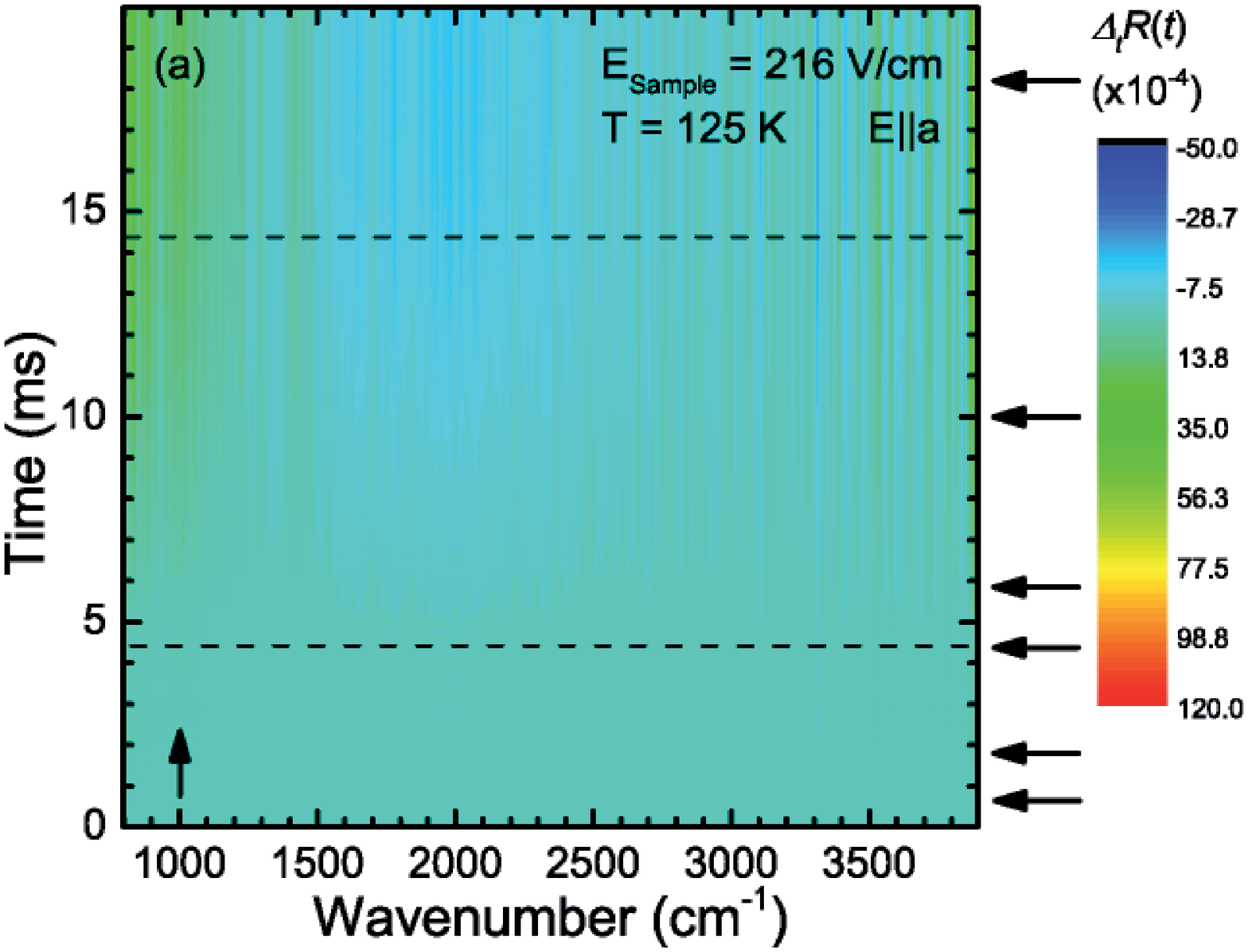}\vspace*{3mm}
       \includegraphics[width=0.8\columnwidth]{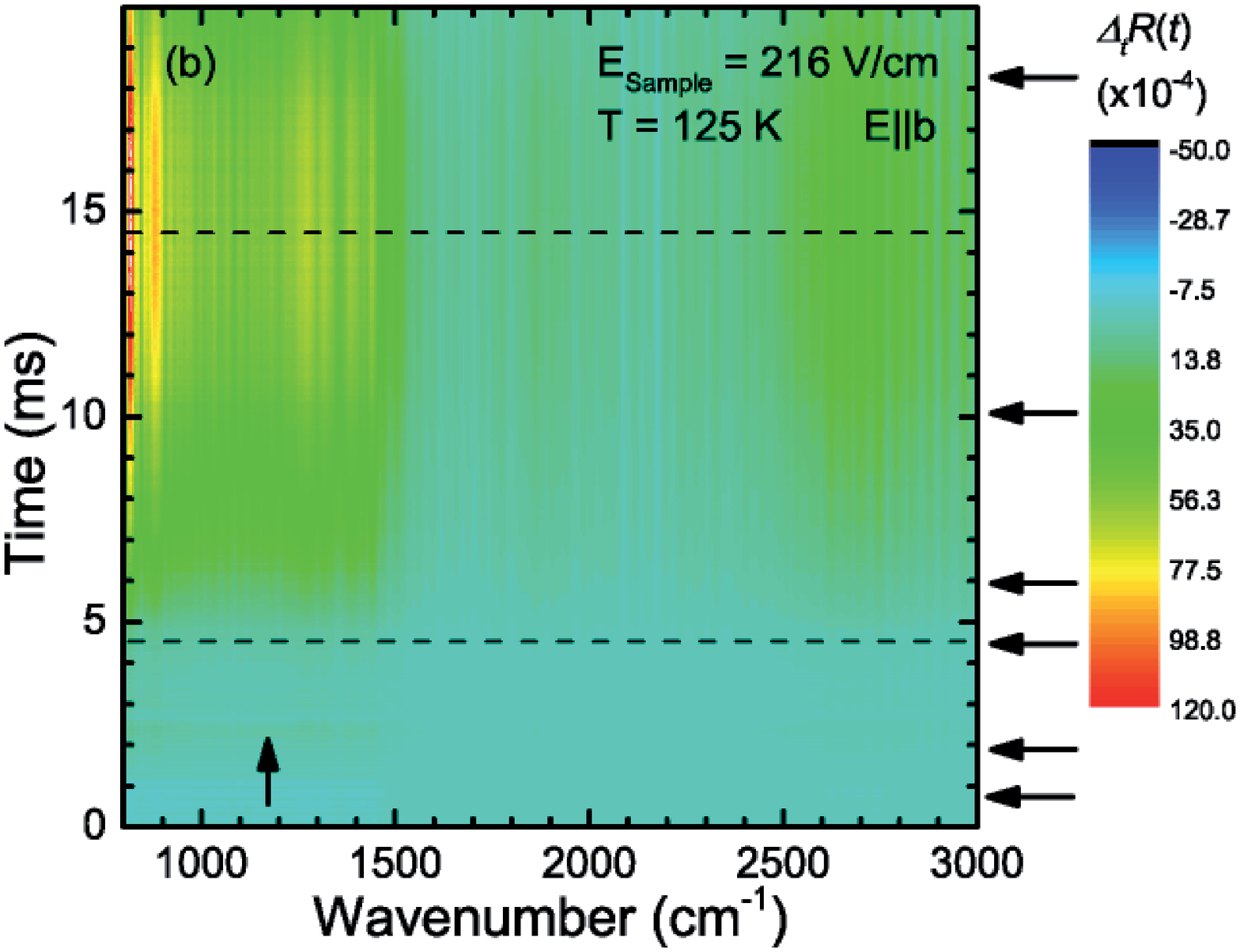}
    \caption{(Color online) Contour plot of the reflectivity change $\Delta_t R(\nu,t)$ along the (a) $a$- and (b) $b$-directions of \aeti\ at $T=125$~K after applying an electric field of $E_{\rm sample}=216$~V/cm along the $a$-axis lasting for 10~ms. The dashed horizontal lines mark the onset and end of the voltage pulse whereas the arrows mark the position where the spectra and the time slices were extracted from the frequency- and time-dependent spectrum and displayed in Fig.~\ref{fig:RefChange125}. The vertical stripes in the spectrum are due to instabilities of the interferometer mirror during the step-scan run.
    \label{fig:DeltaRt125}}
\end{figure}
In Fig.~\ref{fig:DeltaRt125} the time- and frequency-dependent change of the reflectivity $\Delta_t R(\nu,t)$ at $T=125$~K is plotted as recorded for the $a$- and $b$-direction when an electric field of $E_{\rm sample}=216$~V/cm is applied parallel to the crystallographic $a$-axis.
Reflectivity data $R(\nu,t)$ in the mid-infrared range are acquired during a period of 20~ms for both directions.
After 1~ms the voltage pulse is applied to the sample, but it takes another few milliseconds
before the reflectivity changes appreciably.
Only 4~ms after the voltage pulse has started, $R(\nu, t)$ increases for $\nu < 1500$~\cm,
which is accompanied by a drop of the signal between 1500 and 2500~\cm. Qualitatively the behavior is similar for both directions.
After the voltage pulse has stopped at $t=11$~ms, the signal keeps increasing almost to the end of the measurement window. The influence of the electric field on the material exceeds the pulse duration; any modification remains longer than the recorded time frame.

\begin{figure}
    \centering
       \includegraphics[width=1\columnwidth]{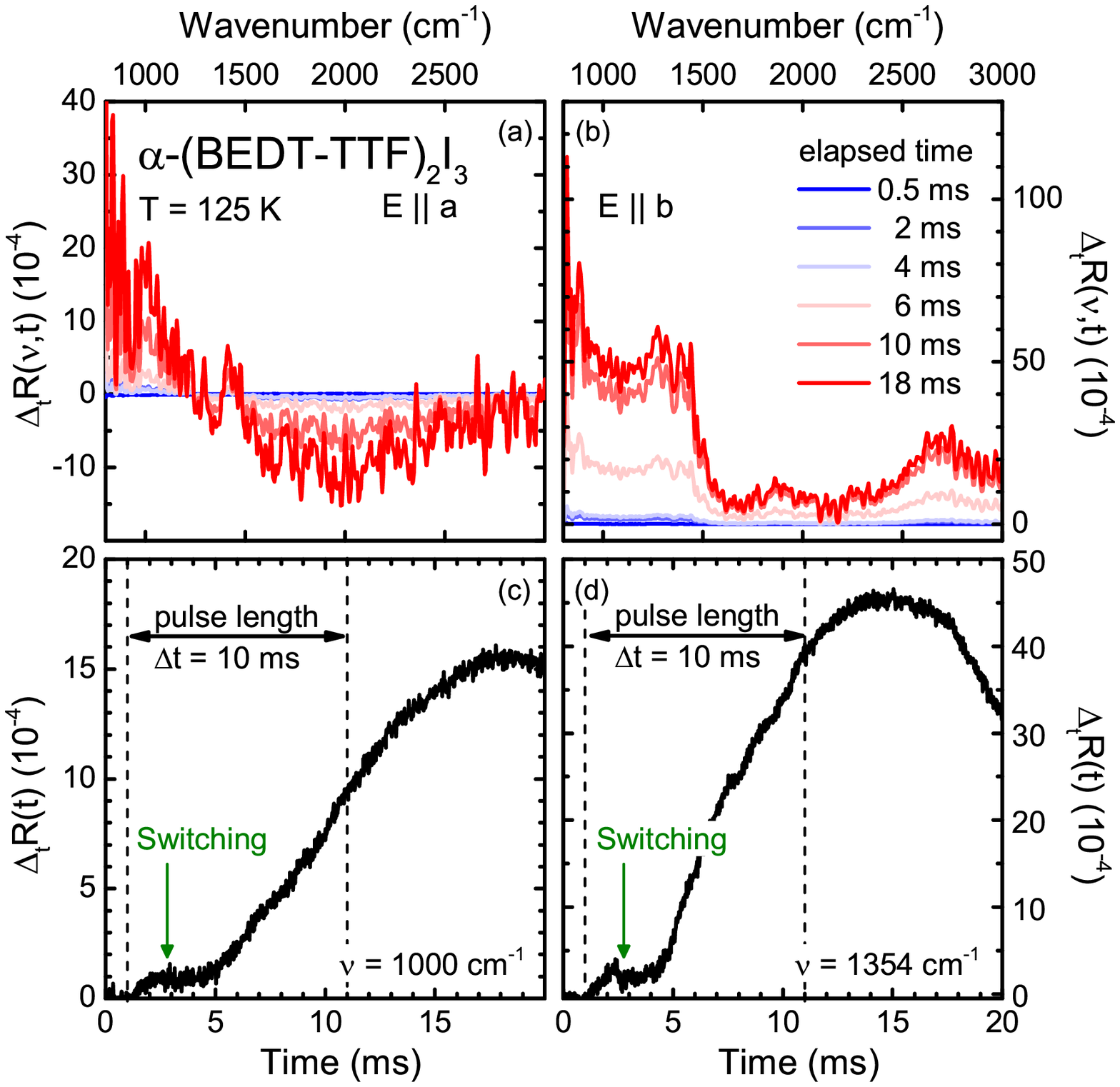}
    \caption{(Color online) (a,b)~Variation of the reflectivity $\Delta_t R(\nu,t)$ spectra of \aeti\ for different times at $T=125$~K. The data in the frequency range from $\nu=800$ to 3000~\cm\ are extracted from Fig.~\ref{fig:DeltaRt125} for both polarizations (a) $E\parallel a$ and (b) $E\parallel b$. The lower panels display the time profile of the reflectivity $R(\nu,t)$ taken at selected frequencies (c) $\nu=1000$~\cm\ and (d) 1354~\cm\ between 0~ms and 20~ms. When the electric field applied, $R(\nu,t)$ increases only slightly then and remains constant in the high-conducting state.
    With a delay of a few milliseconds the signal rises linearly until it saturates after the voltage pulse is switched off; the reflectivity then  decays linearly. The vertical dotted lines mark the start and end point of the voltage pulse.
    \label{fig:RefChange125}}
\end{figure}

In order to analyze the time-dependence of the reflectivity $R(\nu,t)$ in more detail,
we take spectral and time slices at the position marked in Fig.~\ref{fig:DeltaRt125} by black arrows.
In panels (a) and (b) of Fig.~\ref{fig:RefChange125} the evolution of the $\Delta_t R(\nu,t)$ spectra is shown for six different points in time, ranging from  $t=0.5$ to 18~ms. Before the pulse is applied (at $t=0.5$~ms), the reflectivity is steady.
Immediately following the applied field, the reflectivity rises, but at 2~ms $\Delta_t R(\nu,t)$
comes to a halt at a minuscule level where it stays until $t=4$~ms. From our time-resolved transport measurements presented in
Fig.~\ref{fig:ContourResistance}, we can extract that at $T=125$~K the electrically-induced switching occurs 1.7~ms after the field is applied. Thus we cannot relate the phase transition detected in the resistivity drop with the change of the reflectivity at 5~ms. We conclude two different effects being responsible for these observations. By taking a closer look at the spectral shape of $\Delta R(\nu,t)$ after 5~ms and compare them with the reflectivity difference $\Delta_T R$ presented in Fig.~\ref{fig:DeltaRT},
we immediately recognize the similarity in
the difference spectra $\Delta_T R(\nu, T)$ between $T=137$ and 80~K. In other words, the electric pulse transforms the crystal into the metallic phase after 4~ms.

These observations and conclusions are confirmed when we analyze the time-dependent behavior for two selected frequencies, $\nu= 1000$ and 1354~\cm, plotted in Fig.~\ref{fig:RefChange125}(c) and (d).
The signal starts to rise slightly with the onset of the voltage pulse, but it soon levels off.  The electrically-induced switching point at $t=2.7$~ms is not reflected in the $\Delta_t R(t)$ spectra. Only 4~ms after the voltage is applied, the reflectivity starts to rise linearly. This continues well beyond the endpoint of the pulse; $\Delta_t R(t)$ saturates at about 15~ms and then decreases linearly.
With slight variations the overall behavior is similar for both polarization directions.

Several peculiarities call for further discussion:\\
\indent
(1) We suggest that in the first 4~ms of the voltage pulse, a fraction of the sample is transferred into a metallic phase. This picture is supported by spatially-resolved Raman measurements of a current-induced phase transition under steady-state conditions.\cite{Mori09}  Mori {\it et al.} observed a modification of the spectral intensity of a emv-coupled mode of the BEDT-TTF molecules, which they ascribed to the variation of the electronic background and therefore to an electrically-induced metallic state. In contrast, a temperature-sensitive vibrational mode of the I$_3^-$ anions does not reveal any modification. Note, however, that measurements under steady current flow are likely to be subject to heating.\cite{Zimmers13}  Since we cannot correlate the observed switching from a low- to high-conducting state to the temporal dynamics of the sample resistivity, we interpret the change of reflectivity by a thermally-induced phase transition of a few spatially limited areas. The electrical current is affected by some inhomogeneously distributed impurities and cracks within the sample leading to local heating.

(2) It is a remarkable fact that the sample is already in the highly conducting (metallic-like) state according to the transport measurements, but the reflectivity changes only after $t=5$~ms and grows further until 15~ms, although the pulse is already switched off at $t=11$~ms. We suggest that the system is in a highly-conducting state due to an excited electronic system. These  ``excited'' or ``hot'' charge carriers couple to the lattice subsystem to which the energy is eventually transferred with a certain delay time. This leads to a heating-up of the sample across the phase transition causing the reflectivity change. Calculations of the heating effects confirm this picture.\cite{SM}  Here we assumed that the entire electric energy provided by the voltage pulse is completely converted into heat and the sample temperature exceeds $T_{\rm CO}$.

(3) The excited charge carriers relax back to their initial ground state
by releasing the energy to the crystal lattice. $\Delta_t R(\nu,t)$ saturates when the energy transferred to the phonon bath is balanced by the outflow to the heat sink of the sample holder. Eventually all energy is stored in the lattice subsystem; without any influx the temperature rises leading to a linear drop of the reflected signal. This picture consistently  explains also our electric pump-probe experiments\cite{Ivek12} where the resistivity recovers in an exponential manner within a few milliseconds after the highly conducting state was initialized by a 3~ms long voltage pulse.

(4) The linear decrease in $\Delta_t R(\nu,t)$ after 17~ms can be quantitatively described by Newton's law of cooling $Q_{\rm cool} = -\lambda_{\rm therm}(T_{L}-T_0)$, with
$\lambda_{\rm therm}$ is the thermal conductivity, $T_0$ denoting the environment temperature, and $T_L$ the temperature of the heated lattice. Due to the elevated temperature thermal radiation was neglected as a relevant factor.

\subsubsection{Well below the metal-insulator transition: $T=80$~{\rm K}}
\begin{figure}[b]
    \centering
       \includegraphics[width=0.8\columnwidth]{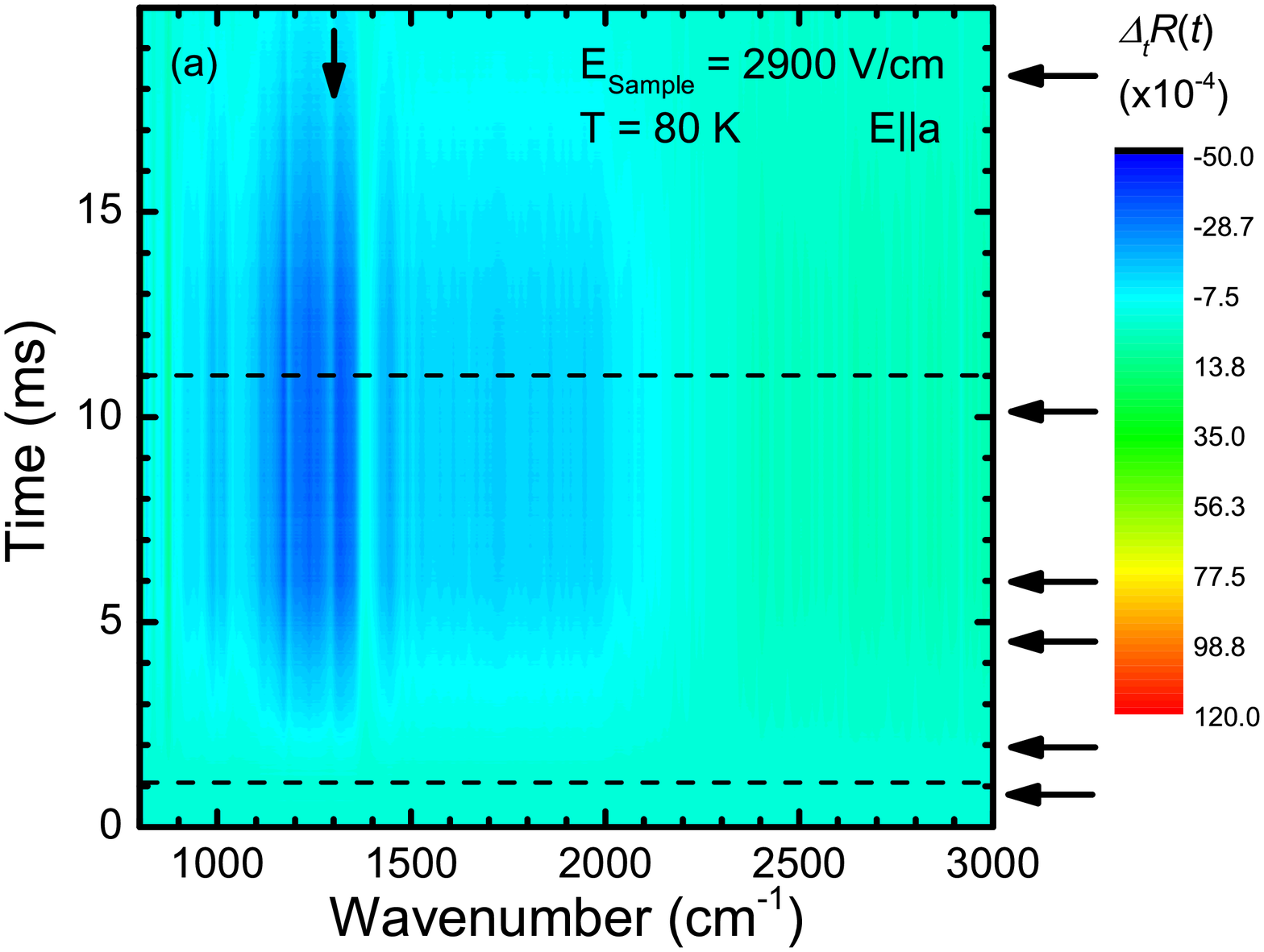}\vspace*{3mm}
       \includegraphics[width=0.8\columnwidth]{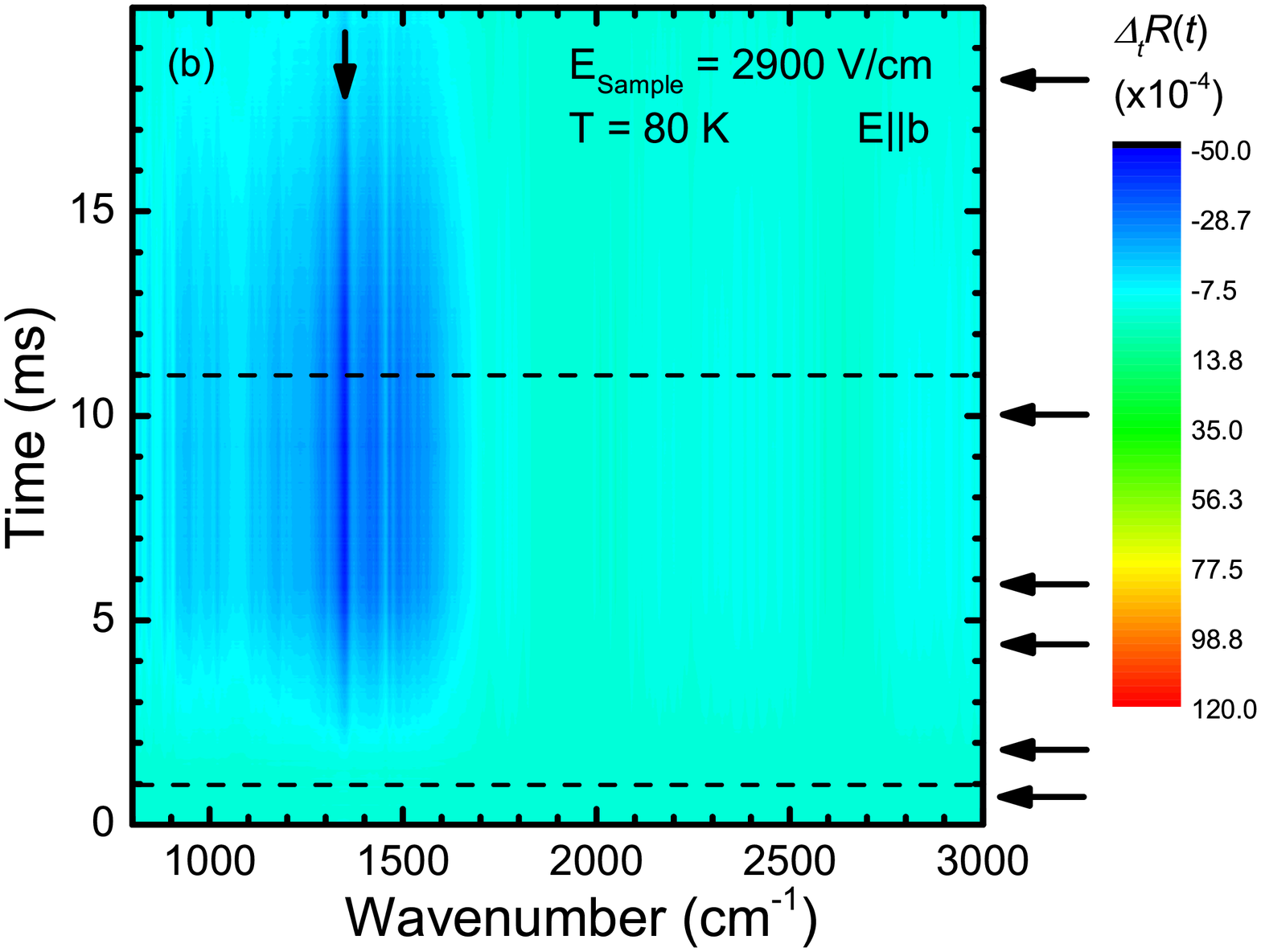}
    \caption{(Color online) Contour plot of the reflectivity change $\Delta_t R(\nu,t)$ of \aeti\ for both polarizations at $T=80$~K  after applying an electric field of $E_{\rm sample}=2900$~V/cm along the $a$-axis lasting for 10~ms. The same color code is used as in the previous Fig.~\ref{fig:DeltaRt125}. The dashed horizontal lines mark the onset and end of the voltage pulse whereas the arrows mark the position where the spectra and the time slices were extracted from the frequency- and time-dependent spectrum and displayed in Fig.~\ref{fig:RefChange080}.
    \label{fig:DeltaRt080}}
\end{figure}
Now, we consider the switching behavior of \aeti\ at $T=80$~K, i.e.\ at temperatures far below $T_{\rm CO}$.
The frequency- and time-dependent spectra $\Delta_t R(\nu,t)$ are displayed in Fig.~\ref{fig:DeltaRt080} for the polarizations $E \parallel a$ and $E \parallel b$,
under an electric field strength $E_{\rm sample}= 2900$~V/cm. The voltage pulse arrives at the sample at time 1~ms and lasts for 10~ms. In both spectra, a change of $\Delta_t R(\nu,t)$ can be recognized right with the onset of the pulse. The important point is, however, that in contrast to the behavior at $T=125$~K depicted in Fig.~\ref{fig:DeltaRt125}, the reflectivity gets smaller and thus $\Delta_t R(\nu,t)$ is mainly negative. It persists until the end of the record time, although it relaxes back about 4~ms after the voltage pulse has been switched off.

\begin{figure}
    \centering
       \includegraphics[width=1\columnwidth]{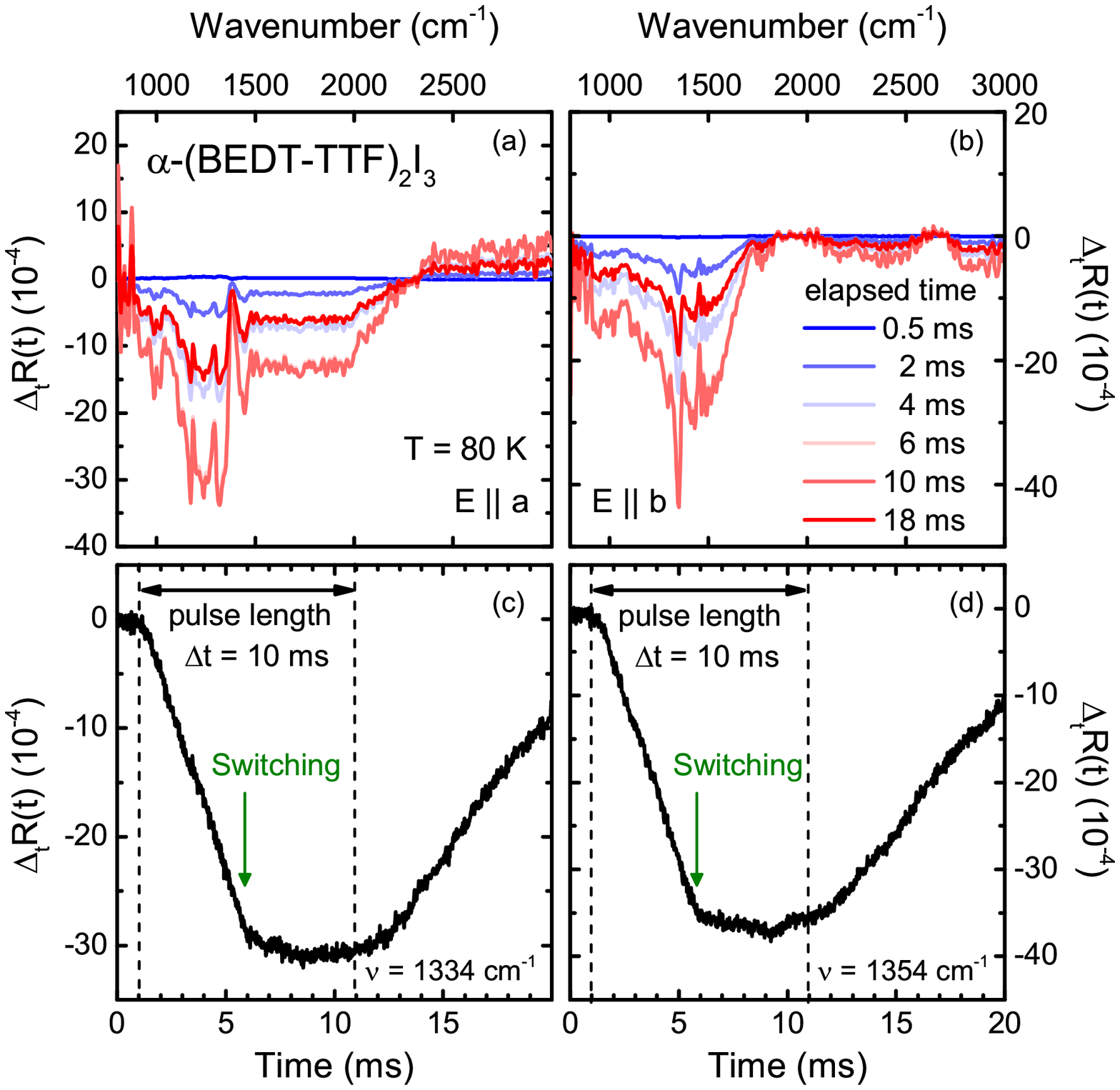}
    \caption{(Color online) (a,b)~Variation of the reflectivity $\Delta_t R(\nu,t)$ spectra of \aeti\ for different times at $T=80$~K. The data in the frequency range from $\nu=800$ to 3000~\cm\ are extracted from Fig.~\ref{fig:DeltaRt080} for both polarizations (a) $E\parallel a$ and (b) $E\parallel b$. The lower panels display the time profile of the reflectivity $R(\nu,t)$ taken at selected frequencies (c)~$\nu=1000$~\cm\ and (d)~$\nu=1354$~\cm\ between 0~ms and 20~ms. When the electric field applied, $R(\nu,t)$ increases only slightly then and remains steady in the high-conducting state.
    With a delay of a few milliseconds the signal rises linearly until it saturates after the voltage pulse is switched off; the reflectivity then  decays linearly. The vertical dotted lines mark the start and end point of the voltage pulse.
    \label{fig:RefChange080}}
\end{figure}
In Fig.~\ref{fig:RefChange080} the frequency-dependent $\Delta_t R(\nu,t)$ is plotted for different points in times, extracted as slices from Fig.~\ref{fig:DeltaRt080}. Immediately after the voltage pulse has arrived at 1~ms, the negative signal increases linearly, saturates at $t\approx 6$~ms and eventually drops after 12.5~ms. For $E\parallel a$ there seems to be one isosbetic point around 2300~\cm, above that frequency $\Delta_t R(\nu,t)$ develops similar, but with opposite, i.e.\ positive sign.
For $E\parallel b$ two frequencies can be identified, $\nu\approx 1900$ and 2650~\cm, with basically no temporal development, but the changes $\Delta_t R(\nu,t)$ are always negative.
The spectra differ significantly from those recorded at $T=125$~K (Fig.~\ref{fig:RefChange125})
and are interpreted as a reduction of the electronic band
and the emv-coupled modes in the case of the $a$- and $b$-direction, respectively.
The comparison with the temperature modulated reflectivity $\Delta_T R(\nu,T)$ shown in
Fig.~\ref{fig:DeltaRT}
reveals a small increase of the sample temperature of only a few Kelvin, for instance,  from $T=80$ to 90~K. Since we do not observe a positive signal below 1000~\cm, the energy gap does not shrink; in excellent agreement with our estimation of an upper limit of the sample temperature caused by Joule heating,\cite{SM}  which predicts a maximum temperature of 96.5~K in the high-conducting state, which is still far below $T_{\rm CO}=135$~K. Therefore, we believe that for our 80~K-experiments no thermally induced phase transition is responsible for the switching from the low-conducing to the high-conducting state.

The time evolution of $\Delta_t R(\nu,t)$ at two frequencies is displayed Fig.~\ref{fig:RefChange080}(c) and (d). The signal decreases in a linear fashion right after the onset of the voltage pulse. Precisely at the resistivity switching from the low-conducting to the high-conducting state at $t=5.5$~ms the signal becomes constant, which implies that the energy flow is balanced between the electron system, the lattice and the environment. When the voltage pulse is turned off, the excited charge carriers
of the electronic system, decays back, the energy is transferred to the lattice system. The infrared signal remains high, exceeding the duration of the applied voltage,  before the lattice system is cooled down linearly in accordance with Newton cooling, as discussed above.

\section{Hot Electron Model}
\label{sec:model}
In order to explain our observations, we suggest a model of ``hot'' electrons or non-equilibrium charge carriers.\cite{Kroll74,Sze07,Yu10} It assumes that the electric power is first stored in the electron system causing an increase of the electron temperature $T_e$. The coupling to the lattice subsystem is responsible for heating up the lattice and to increase the lattice temperature $T_L$. Two types of carriers are assumed with rather different mobilities.
More details are presented in the Supplemental Materials.\cite{SM}

\subsection{Simulation of time dependence}
\label{sec:time_dependence}
The electron temperature $T_e$ can be calculated by considering energy conservation:
\begin{equation}
nC_e\frac{{\rm d}T_e}{{\rm d}t}=P-K \quad ,
\label{eq:HotModel}
\end{equation}
where $P=\sigma(T_{e})E_{\rm sample}^2$ describes the power gained by the electrons and $K=-\lambda_{\rm therm,e}\left(T_{e}-T_{L}\right)$ corresponds to the energy transfer from the electrons to the lattice subsystem\cite{remark2}.
We have solved Eq.~(\ref{eq:HotModel}) numerically by first calculating the right side and subsequently determining $T_e$ from the left side for each time step of the voltage pulse.
The electrical conductivity $\sigma(T_{e})$ is taken from dc measurements plotted in
Fig.~S2;
the load resistance $R_L$ was experimentally selected  between 1~k$\Omega$ and 100~k$\Omega$. In addition we choose an electronic heat capacity $nC_e=0.2~{\rm JK}^{-1}{\rm cm}^{-3}$ and thermal conductivity of the electron system $\lambda_{\rm therm,e}=40~{\rm WK}^{-1}{\rm cm}^{-3}$ in agreement with previous investigations,
where a similar two-state model was applied to other low-dimensional organic conductors.\cite{Mori09,Ozawa09}

\begin{figure}
    \centering
       \includegraphics[width=1\columnwidth]{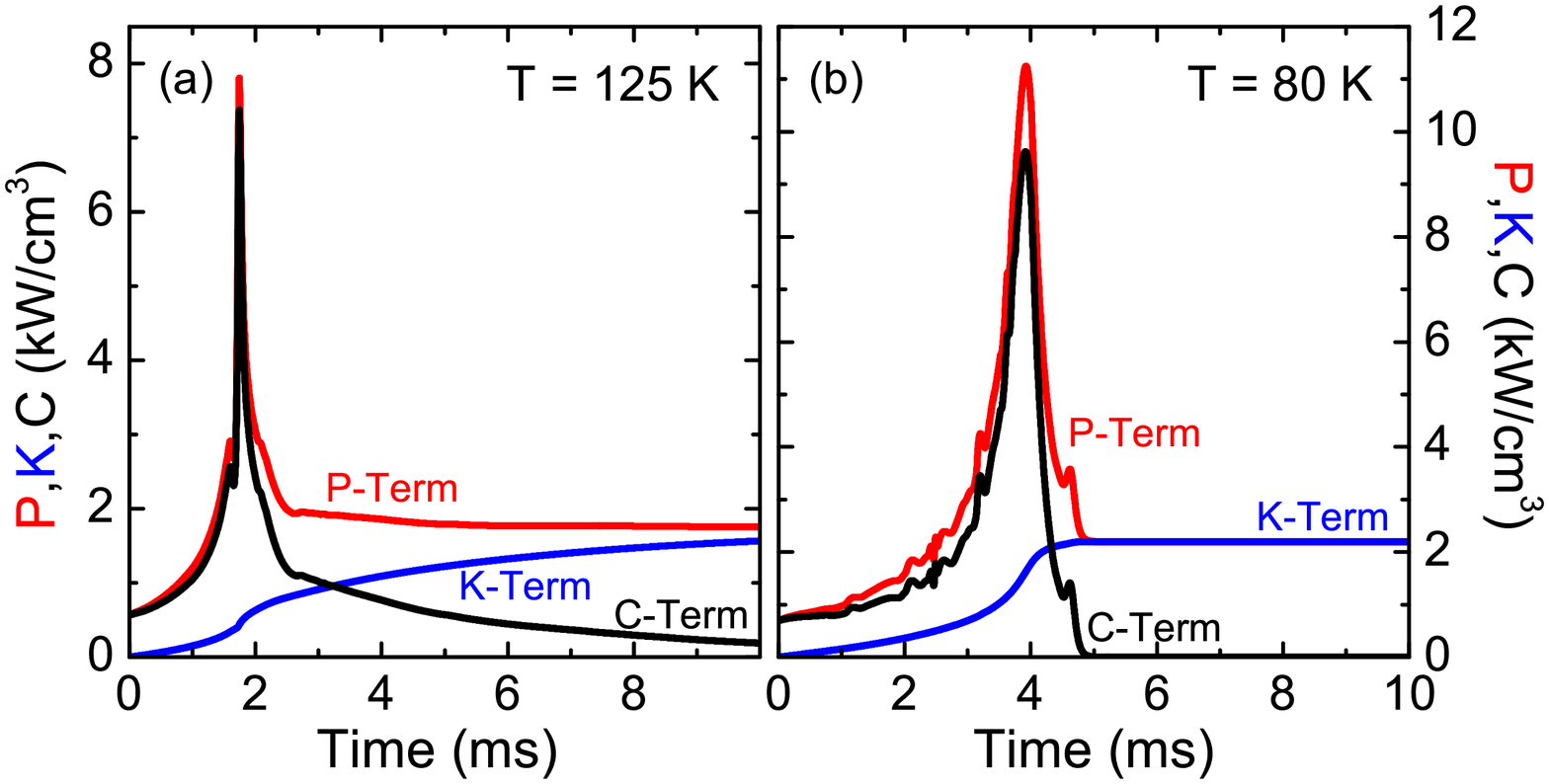}
    \caption{(Color online) Time evolution of the input power $P$ (red), the cooling power $K$ (blue) and and the power absorbed by the electronic system (black) calculated for the temperature (a) $T=125$~K and (b) $T=80$~K, where the applied electric field was $E_{\rm sample}=250$ and 3000~V/cm, respectively.
    \label{fig:Power}}
\end{figure}
In Fig.~\ref{fig:Power} the time dependences of the $C$-, $P$-, and $K$-terms are plotted for $T=125$
and 80~K. The electric power is transferred to the electronic system right from the beginning of the voltage pulse.
The cooling term, however, increases only slowly for the elevated temperature $T=125$~K. Before the switching point, the electric power as well as the stored energy in the electric system diverge
and drop afterwards to a fixed value. The $C$-term decreases slowly to zero since the cooling term approaches the electric power term $P$ initializing the steady-state.
At low temperatures, $T=80$~K, the power term behaves similarly, only the cooling term $K$ immediately  after the resistivity switching reaches the value of the electric input $P$-term; correspondingly
no energy is then transferred to the electronic system.

\begin{figure}
    \centering
       \includegraphics[width=0.8\columnwidth]{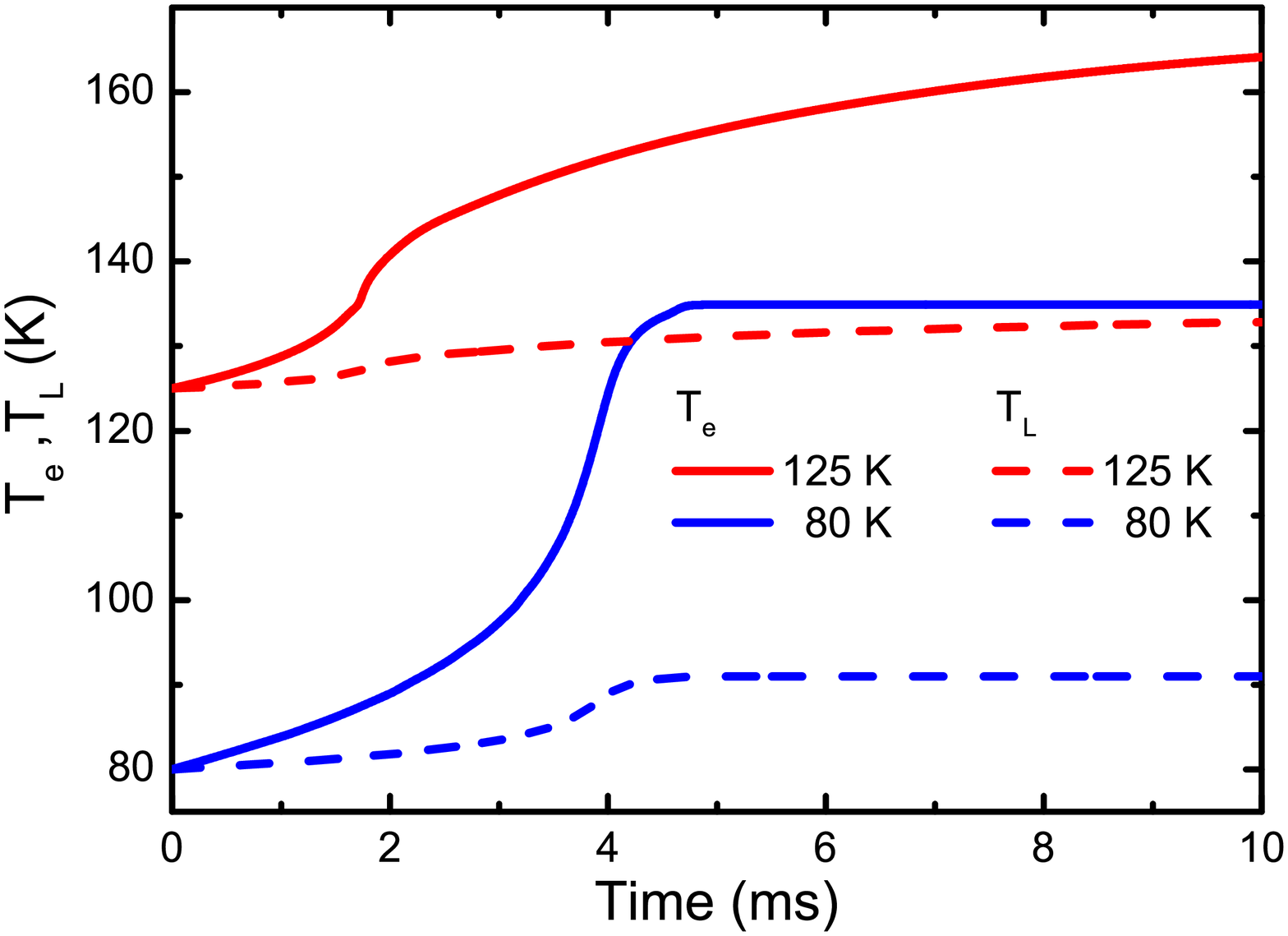}
    \caption{(Color online) The electron temperature $T_e$ (solid lines) and the lattice temperature $T_L$ (dashed lines) plotted as a function of time for the two experimental temperatures $T=125$~K (red) and $T=80$~K (blue) by applying an external electric field of $E_{\rm sample}= 250$ and 3000~V/cm respectively.
    \label{fig:Temperature}}
\end{figure}

The time evolution of the electron temperature $T_e$ is displayed in Fig.~\ref{fig:Temperature}
for the two experiments at different temperatures.
When starting at $T=125$~K, the electron temperature increases more than linearly,
then reveals a step-like feature at the phase transition around 2~ms and finally increases sub-linearly up to 160~K at the end of the pulse.
For the low-temperature experiment ($T=80$~K) the temperature rises steeply up to $T=134$~K and stays constant; right below the transition temperature $T_{\rm CO}$.

The total heat capacity\cite{Fortune91} of \aeti\ crystals below $T_{\rm CO}$, $C_{\rm tot}=0.5~{\rm JK}^{-1}{\rm g}^{-1}$, is more than five times larger than the values for the electron system $C_e=
0.09~{\rm JK}^{-1}{\rm g}^{-1}$ used in our simulations.
This agrees well with the fact, that phonon typically dominate above 10~K due to the $T^3$ power-law of the lattice contribution compared to the linear dependence of the electronic heat capacity.
Assuming that both contributions are independent and add up,  $C_{\rm total}=C_{e}+C_{L}\approx 5C_{e}$,
energy conservation yields:\cite{Mori09} $nC_{e}\left(T_{e}-T_{L}\right)= nC_{L}\left(T_{L}-T_{0}\right)$ and thus
\begin{equation}
\frac{C_{e}}{C_{L}} = \frac{(T_{L}-T_0)}{(T_{e}-T_{L})}=\frac{1}{4}
\label{eq:Lattice}
\end{equation}
with the initial environmental temperature $T_0$; here $T_e$ is the effective electron temperature and $T_L$ denotes the lattice temperature due to heating.

Using Eq.~(\ref{eq:Lattice}) we estimate the time dependence of the lattice temperature of \aeti\ for $T=125$ and 80~K and display the results in Fig.~\ref{fig:Temperature}. Trailing the electron temperature, the lattice temperature increases only slightly from $T=125$ to 132~K within the pulse duration of 10~ms; hence it stays below the charge-ordering temperature of $T_{\rm CO}=135$~K.
Of course, we cannot exclude that due to some scattering centers $T_L$ can locally exceed the transition temperature.
We now understand, why the reflectivity $\Delta_t R(\nu,t)$ still increases after switching to the high-conducting state; the lattice temperature $T_L$ keeps rising well after the system has switched.
From Fig.~\ref{fig:RefChange125} we clearly see a metallic transition in the 125~K run,
but at $T=180$~K the reflectivity variation indicates only a small increase of $T_L$ of less than 10~K. This is exactly what our simulation reproduce: the lattice temperature first increases  but then saturates as the system switches from the low-conducting to the high-conducting state. This explains why $\Delta_t R(\nu,t)$ remains constant in time until the pulse is switched off.

\subsection{Simulation of field dependence}
\label{sec:field_dependence}
\begin{figure}
    \centering
       \includegraphics[width=0.8\columnwidth]{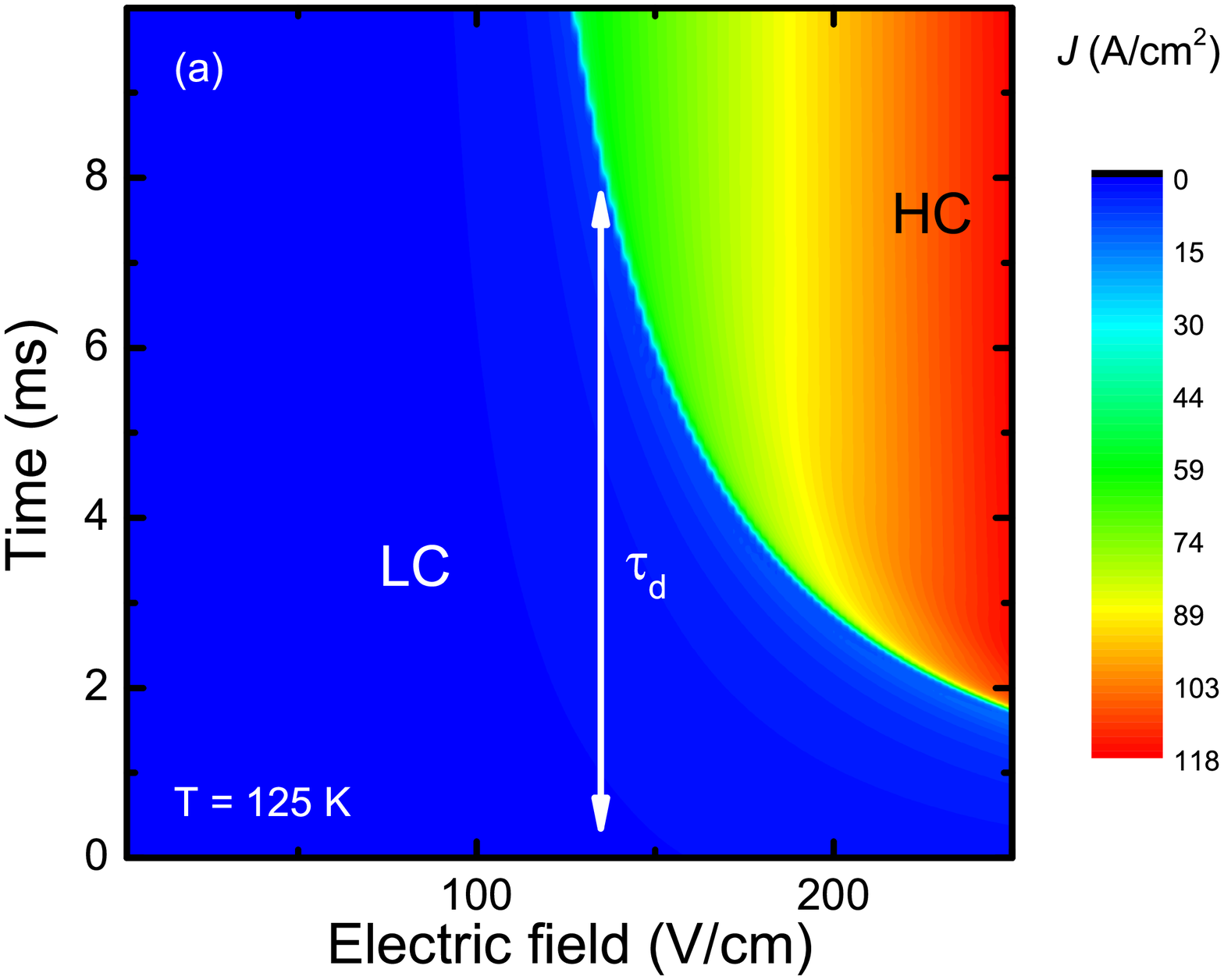}\vspace*{3mm}
       \includegraphics[width=0.8\columnwidth]{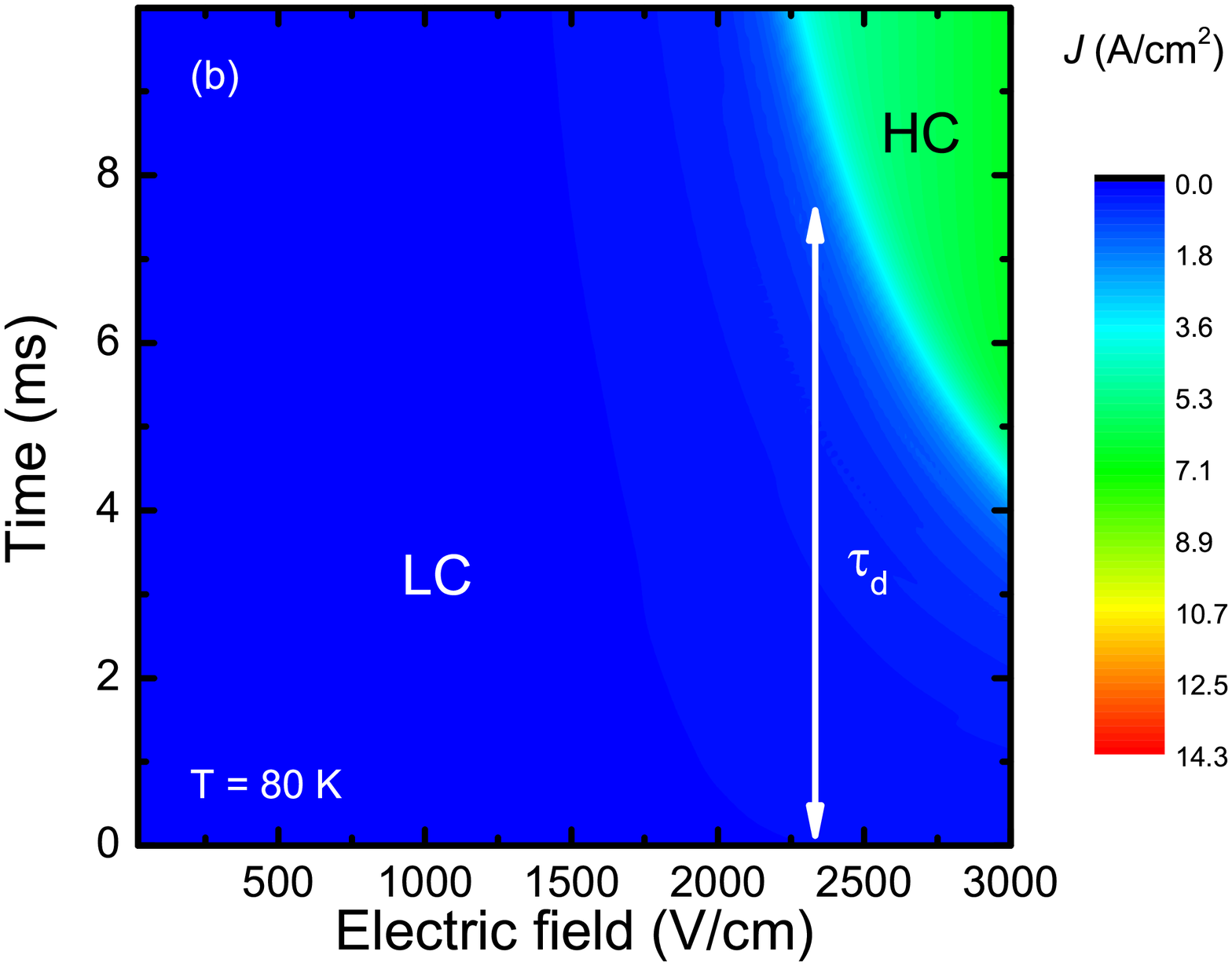}
    \caption{(Color online) Variation of the current density $J(E,t)$ with  increasing electric field $E_{\rm sample}$ applied to the sample. Simulation for the temperature (a) $T=125$~K and (b) $T=80$~K; note the different scales. The arrows indicate the delay time $\tau_d$. The low-conducting state (LC, blue area) is characterized by a small current density while in the high-conducting part (HC, green-red) the current density is extremely high.
    \label{fig:Simulations}}
\end{figure}

Up to now, only the time-dependent behavior of the power and the different temperatures were discussed. For a better comparison with our experiments, we also have calculated the current density $J(E,t)$ when the applied electric field is varied. In Fig.~\ref{fig:Simulations} the time-dependence $J(t)$ is plotted for $T=125$ and 80~K when the external electric fields is varied from 0 to 250~V/cm and to 3000~V/cm, respectively.
Note, that $J$ is inversely proportional to the sample resistance $R$, thus
Fig.~\ref{fig:Simulations} can be readily compared with the presentation of our experimental transport results in Fig.~\ref{fig:ContourResistance}.

A very distinct jump of $J(E,t)$ occurs when $E_{\rm sample} > 100$~V/cm providing evidence for the transition from a low-conducting to a high-conducting state; this is in excellent accord with the experimental observations. Also the field-dependent behavior of the delay time $\tau_d$ describes the experimental behavior perfectly.
In the simulation the switching occurs at 2000~V\cm, while our measurements yield $E_{\rm sample}= 2900$~V/cm. The small deviations between simulations and experiments may be due to the
neglect of the contact resistance in our two-point measurement [Fig.~\ref{fig:SetupPulse}(a)];
it influences $E_{\rm sample}$ and its $T$ dependence.
Furthermore, we assumed a temperature-independent specific heat $n C_p$ and thermal conductivity $\lambda_{\rm therm,e}$.
Nevertheless, we can conclude that the two-state model perfectly reflects our experimental findings and yields an excellent qualitative and quantitative description.

\begin{figure}
    \centering
       \includegraphics[width=0.8\columnwidth]{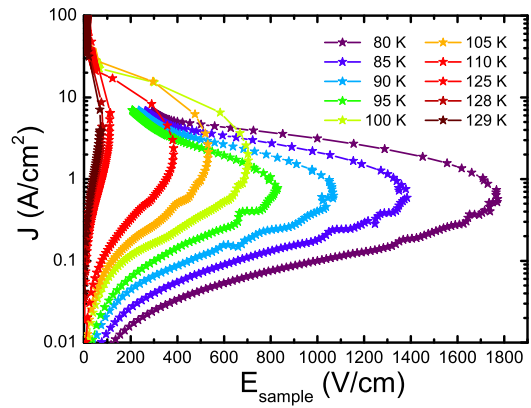}
    \caption{(Color online) Simulated $J$-$E$ characteristic between 0 and 1900~V/cm from temperatures
    $T=130$ to 80~K. All curves exhibit a nonlinear regime with a negative differential resitance. On cooling, the threshold current decreases while the threshold electric field rises.
    \label{fig:Jnonlinear}}
\end{figure}
In a next step, the $J$-$E$ characteristic is calculated to investigate the nonlinear conductivity and the negative differential resistance (NDR). The results are depicted in Fig.~\ref{fig:Jnonlinear} for temperatures between $T=80$ and 130~K. The overall behavior is very similar to the data plotted in Fig.~\ref{fig:Current_log} resembling an S-shaped current-voltage curve. All temperatures reveal a linear regime at low fields, they exhibit a turning point which marks the threshold field $E_{\rm thesh}$ and current density $J_{\rm thesh}$. All $J$-$E$ curves cross over in the NDR regime.
Above a certain current density, the field increases again but much steeper than for lower electric fields, indicating that the high-field conduction state is different from the initial state at low fields.
Small deviations from the experimental findings can be noted, such as
slightly higher current density after the NDR compared to the experiment, or the higher fields for which the curves turn. We attribute these to the unknown contact resistance in the experiment which is in series to the sample and load resistance and affects $E_{\rm sample}$ and $J$, directly. Nevertheless, the detailed agreement of the simulation with the experiment is remarkable.

\begin{figure}
    \centering
       \includegraphics[width=1\columnwidth]{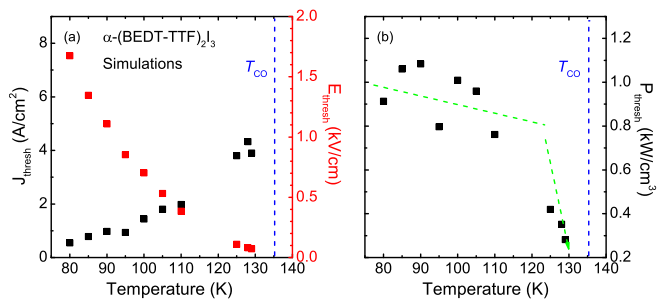}
    \caption{(Color online) (a) Numerical simulations of the threshold electric field (red) and current density (black) as a function of temperature below $T_{\rm CO}$ using parameters for \aeti. (b) Temperature dependence of the threshold power $P_{\rm thesh}$ determined from the product of $E_{\rm thesh}$ and $J_{\rm thresh}$. The vertical dotted line marks the phase transition temperature of $T_{\rm CO}=135$~K.
    \label{fig:Threshold_sim}}
\end{figure}
Following our analysis of the experimental data in Sec.~\ref{sec:transport_studies},
we have extracted the threshold current $J_{\rm thresh}$ and the threshold field $E_{\rm thresh}$ from the $J$-$E$ curves in Fig.~\ref{fig:Jnonlinear} and plotted them as a function of temperature in Fig.~\ref{fig:Threshold_sim}. The results of our simulations resemble more-or-less closely the experimental data displayed in Fig.~\ref{fig:Threshold_exp}.
The observed threshold field $E_{\rm thresh}(T)$ was well described by a linear increase
as the temperature is reduced below $T_{\rm CO}$, our simulation, however, rise in a quadratic manner.
This may be attributed to the decrease of the occupation number of the conduction band on cooling.
Moreover, the increase of $E_{\rm thesh}$ at low temperatures can also be attributed by thermal depletion of the conduction band upon cooling.

The calculated current density, on the other hand, decays linearly on cooling, while the experiments
seem to flatten at low temperatures. However, the threshold power density $P_{\rm thresh}(T)$ increases steeply close to the phase transition and crosses over in regime with a smaller slope, which resembles the development of the energy gap.

We have seen that the two-state model excellently describes the electrically-induced phase transition and explains qualitatively and, impressively, also the quantitative behavior of the experimental values. It does not only support the time-resolved transport measurements, but also explains in detail the observations made in the time-dependent reflectivity study. It confirms that we are dealing with an electronic system linearly coupled to the crystal lattice. Furthermore, we made a precise estimation of the lattice temperature $T_L$ which is different from the electronic system and agrees nicely with estimate sample temperature derived from the comparison of time-dependent reflectivity with the steady-state spectra.

\subsection{Field-induced Dirac-like electrons}
We have demonstrated that switching into a high-con\-duct\-ing state is caused by excitations of charge carriers into a state with a high mobility.
Here, the question arises whether the enhanced mobility is somehow related to the linear band dispersion found in \aeti. The tilted Dirac cone, i.e.\ the touching of the bands at the Fermi energy, is only predicted for high pressure. Nevertheless, band structure calculations (Fig.~S4)
yield that the linear dispersion exists also in the insulating phase.\cite{SM,Peterseim16e} Furthermore, the temperature-dependence of the resistance under pressure normalized to the room-temperature values  exhibit exactly the same slope as the resistance at ambient pressure.\cite{Schwenk85,Tajima06} Since Dirac fermions are made responsible for the conductivity behavior under pressure, we propose, that they also contribute to the metallic state at ambient conditions, excited by a strong electric field.
The coexistence of two different types of charge carriers with two orders of magnitude different mobility was previously suggested based on magnetotransport\cite{Monteverde13}  and pressure-dependent optical experiments\cite{Beyer16} in \aeti.
We propose that also the electrons excited in the conduction band can act either as normal massive electrons or as massless Dirac-like electrons with very high mobility. This picture provides a consistent explanation of the results presented here.

\section{Conclusions}
We have reported comprehensive time-resolved investigations of the electrodynamics
at the metal-insulator transition in \aeti.
In the charge-ordered state below $T_{\rm CO}=135$~K,
the time- and field-dependent charge transport was measured
by applying a voltage pulses along the $a$-direction.
We found an electrically-induced resistivity switching from a low-conducting to a high-conducting state,
leading to an S-like shape in the $J$-$E$ curve. In a regime of negative differential resistance, the electric field across the sample drops drastically above a certain threshold, which increases upon cooling.

Measuring the polarization-dependent infrared reflectivity after a strong electric field was applied,
allows us to study the transient optical properties in a wide temperature and field range.
From the spectral signature at $T=125$~K , we can identify a field-induced metallic state similar to the high-temperature conducting phase.
In contrast, at temperatures well into the insulating state ($T=80$~K)
the reflectivity variations are completely different and cannot be explained by Joule heating.
We suggest the creation of an electronically driven high-conducting state in \aeti.
These field-excited charge carriers exhibit an extremely  high mobility
known from the Dirac-like electrons present under high pressure. Thus we propose that the electrons excited in the conduction band can act either as normal massive electrons or as massless Dirac-like electrons with very high mobility.

This generation of hot charge carriers by the electric field is then simulated with a two-state model.
Only based on energy conservation and known material parameters,
our numerical simulations can -- qualitatively as well as quantitatively -- reproduced
the experimentally observed properties, such as the temporal behavior of the resistivity as a function of the electric field, the negative differential resistivity, and the different threshold values. By defining an effective electron temperature $T_e$, we estimate the lattice temperature $T_L$ close to the measured sample temperature.
For all parameters analyzed, the quantitative agreement between model and experiment is excellent.

\acknowledgments
We would like to thank
R. Beyer, E. Rose, D. Wu and S. Zapf for useful discussions as well
as G. Untereiner for technical support. Funding by the Deutsche
Forschungsgemeinschaft (DFG) and Deutscher Akademischer
Austauschdienst (DAAD) is acknowledged.


\begin{thebibliography}{62}
\bibitem[{rem()}]{remark2}
\bibinfo{note}{For our calculations we set the lattice temperature $T_L$ equal
  to the environment temperature $T_0$. The last term corresponds to the
  experimental situation described by Fig.~\protect\ref{fig:RefChange080}(b)
  and (d): the reflectivity does not change and correspondingly the sample has
  a constant temperature. We have assumed a linear Newton cooling.}
\end{thebibliography}

\end{document}